\documentclass[aps,prb,reprint]{revtex4-1}
\usepackage[dvips]{graphicx}
\usepackage{amsmath}
\usepackage{color}

\usepackage{graphicx} 
\usepackage{epstopdf}

\begin{document}

\title{Exceptional and Anisotropic Transport Properties of Photocarriers in Black Phosphorus}

\author{Jiaqi He$^{1,2}$, Dawei He$^{1}$, Yongsheng Wang$^{1,*}$ Qiannan Cui$^{2}$, Matthew Z. Bellus$^{2}$, Hsin-Ying Chiu$^{2}$, \& Hui Zhao$^{2,*}$}

\affiliation{$^1$ Institute of Optoelectronic Technology, Beijing Jiaotong University, Beijing 100044, China\\
$^2$Department of Physics and Astronomy, The University of Kansas, Lawrence, Kansas 66045, USA}

\date{\today}

\begin{abstract}

We show that black phosphorus has room-temperature charge mobilities on the order of 10$^4$ cm$^2$V$^{-1}$s$^{-1}$, which are about one order of magnitude larger than silicon. We also demonstrate strong anisotropic transport in black phosphorus, where the mobilities along the armchair direction are about one order of magnitude larger than zigzag direction. A photocarrier lifetime as long as 100 ps is also determined. These results illustrate that black phosphorus is a promising candidate for future electronic and optoelectronic applications.

\end{abstract}

\maketitle

The rapid development of silicon-based integrated circuit technology over the past several decades relied on shrinking the channel length of the transistors. This approach is facing a severe challenge posted by several fundamental limits. Hence, recently, significant efforts have been devoted to finding new materials for post-silicon electronic technology. One promising candidate, graphene, has shown room-temperature charge mobilities two orders of magnitude higher than silicon\cite{nm6183,rmp81109}. However, the lack of a band gap limits its application as a channel material in logic devices. Since 2012, transition metal dichalcogenides, such as MoS$_2$, have been extensively studied\cite{nn7699}. These monolayer semiconductors have sizable band gaps\cite{l105136805,nl101271} and possess several novel optical and spin/valley-related properties\cite{l108196802}. However, their charge mobilities are still lower than desired. 

Since 2014, two-dimensional black phosphorus (BP), also known as phosphorene, has emerged as a new promising material to provide both a sizable band gap and high charge mobilities. Few-layer and monolayer BP can be fabricated from bulk crystals by mechanical\cite{acsnano84033} and liquid\cite{cc5013338} exfoliation. Although bulk BP has a relatively small band gap\cite{jap341853} of 0.3 eV, several studies\cite{nl145733,nl146400,acsnano84033,acsnano89590} have shown that the band gap of BP thin films depends strongly on the thickness when approaching the monolayer limit - a feature similar to other two-dimensional materials: Charge transfer characteristics of field-effect transistors indicated a band gap of 1.0 eV in monolayers\cite{nl145733}; while a 2.0-eV monolayer band gap was suggested by a scanning tunneling microscopy measurement on the surface of a bulk BP\cite{nl146400}. Photoluminescence peaks from excitons were observed at 1.45 eV in monolayer BP\cite{acsnano84033}, and evolves from 1.3 to 0.9 eV when the number of atomic layers are changed from two to five\cite{acsnano89590}. Without knowing the exciton binding energies, these numbers give the lower limits of the band gap. Although more studies are still necessary to precisely determine the band gaps, the fact that few-layer BP possesses band gaps large enough for logic applications has been established. On transport properties, theoretical studies indicated that monolayer and few-layer BP possess high room-temperature mobilities\cite{nc54475} on the order of 10$^4$ cm$^2$V$^{-1}$s$^{-1}$ and can outperform silicon and MoS$_2$ in ballistic devices\cite{ieeeedl35963,ieeeted613871}. Experimentally, field-effect transistors of multilayer BP have been fabricated\cite{acsnano84033,acsnano810035,nc54458,nn9372,ieeeedl35795,acsnano811730}. These transistors can have long operation lifetime once passivated by top Al$_2$O$_3$ layers\cite{acsnano811753,nl146964}, and can operate at GHz regime\cite{nl146424}. In optoelectronic applications, high photo-responsitivity\cite{nl143347} and broadband response to both visible and infrared radiations\cite{nl146414} have been demonstrated in BP. Utilizing the ambipolar transport behavior, photovoltaic effect was demonstrated in few-layer BP in a local-gate p-n junction configuration\cite{nc54651}.

Here we report experimental observations of exceptional transport performance of multilayer BP and its high in-plane anisotropy. By resolving spatiotemporal dynamics of photocarriers that are instantaneously injected by a tightly focused ultrashort laser pulse, we obtain room-temperature diffusion coefficients of 1,300 and 80 cm$^2$s$^{-1}$ for photocarriers moving along the armchair and zigzag directions, respectively. These values correspond to ambipolar mobilities on the order of 50,000 and 3,000 cm$^2$V${-1}$s$^{-1}$, respectively. We also find that the photocarrier lifetime in multilayer BP is about 100 ps. These results provide fundamental parameters of BP, and illustrate its great promise for electronic and optoelectronic applications.

Bulk BP crystal has orthorhombic symmetry, as illustrated in Fig. 1{\bf a}, where single layers of phosphorus atoms are stacked by van der Waals force. In each layer, phosphorus atoms form a puckered honeycomb lattice, as shown in Fig. 1{\bf b}. Thin films of BP were mechanically exfoliated from bulk crystals purchased from 2D Semiconductors. An adhesive tape is used to remove layers of BP from the crystal, which are then transferred to a SiO$_2$ (90 nm)/Si substrate. The identified flake is immediately covered by a thin boron nitride layer, which was prepared with the same procedure, in order to prevent its degradation in air. The sample used in this study is shown in Fig. 1{\bf c}. The thicknesses of the BP film and the boron nitride layer are determined to be 16 and 4 nm, respectively, according to atomic force microscopy measurements.  All the measurements were performed in ambient condition. No sample degradation was observed during the entire course of the study.

\begin{figure}
 \centering
  \includegraphics[width = 8.5 cm]{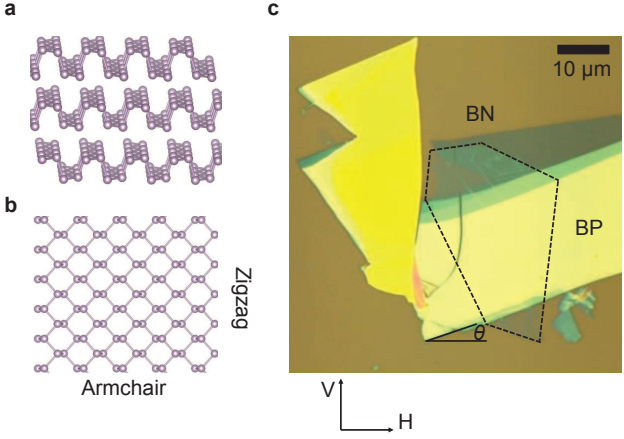}
  \caption{{\bf Lattice structure of black phosphorus and sample image}. {\bf a}, Lattice structure of multilayer black phosphorus. {\bf b}, Top view of a monolayer black phosphorus, with armchair and zigzag directions. {\bf c}, Optical image of the sample studied. The black phosphorus (BP) flake of 16-nm thick is partially covered by a 4-nm boron nitride (BN) flake, indicated by the dashed line. The horizontal (H) and vertical (V) directions in the laboratory frame are labeled. The angle $\theta$ is between H and the long edge of the flake.}
  \label{sample}
\end{figure}

In our transient absorption measurements, a 730-nm and 100-fs laser pulse with a peak fluence of 10 $\mu$J~cm$^{-2}$ injects electron-hole pairs in the sample. The pulse is linearly polarized along the long edge of the flake, as shown in Fig. 1{\bf c}. The dynamics of the injected photocarriers are monitored by measuring the differential reflection of a 810-nm probe pulse that is polarized along the direction perpendicular to the pump polarization. The differential reflection of the probe is defined as the normalized change in the probe reflection induced by the pump, $\Delta R / R_0 \equiv (R-R_0)/R_0$. Here, $R$ and $R_0$ are the reflection of the probe with the presence of the pump pulse or without it, respectively.  As shown in Fig. 2{\bf a}, the differential reflection signal reaches a peak near zero probe delay. After a transient process of about 20 ps, the signal decays exponentially (red line), with a time constant of 100 $\pm$ 5 ps. We attribute this to the carrier lifetime in BP. By repeating the measurement with different values of the pump fluence (and hence different densities of the injected carriers), we confirm that the peak differential reflection signal is proportional to the pump fluence, as shown in the inset of Fig. 2{\bf a}. 

\begin{figure}
 \centering
  \includegraphics[width = 8.5 cm]{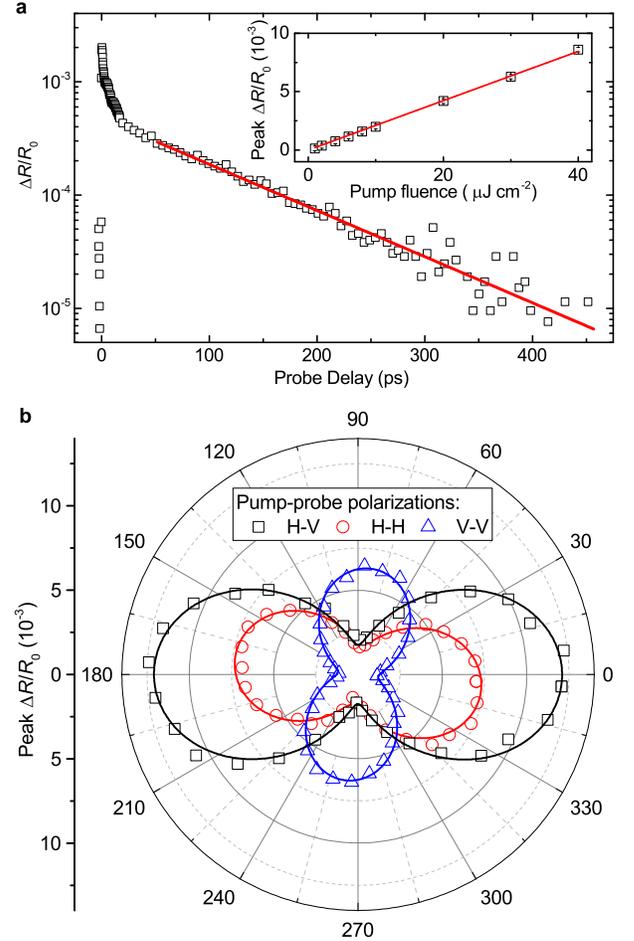}
  \caption{{\bf Time-resolved differential reflection signal from black phosphorus and its angular dependence}. {\bf a}, Differential reflection ($\Delta R / R_0$) measured from the sample with a 730-nm pump and a 810-nm probe pulses. The red line indicates a single-exponential fit with a time constant of 100 $\pm$ 5 ps. The inset shows linear relation between the peak signal and the pump fluence. {\bf b}, The peak differential reflection signal as a function of $\theta$, the angle between the long edge of the flake and horizontal direction in laboratory frame. The black squares, red circles, and blue triangles are measured with the pump-probe polarization configurations of horizontal-vertical (H-V), horizontal-horizontal (H-H), and vertical-vertical (V-V), respectively.}
  \label{rotation}
\end{figure}

The lattice structure of BP shown in Fig. 1{\bf a} dictates that its in-plane properties are anisotropic, since the effective masses of the electrons and holes depend strongly on the crystalline direction\cite{nc54475,jap116214505,nl142884,b90115439}. Indeed, the anisotropic optical responses of BP have been predicted\cite{nc54475,b90075434} and experimentally confirmed by polarization-dependent Raman scattering\cite{acsnano89590,nc54458} and absorption\cite{nc54458}. To study the anisotropic transient absorption of BP, we measure the differential reflection signal with different pump-probe polarization configurations as we rotate the sample with respect to the polarization directions. The squares, circles, and triangles in Fig. 2{\bf b} show the peak signal of differential reflection as a function of the angle ($\theta$) between the long edge of the flake and the horizontal direction in laboratory frame (Fig. 1{\bf c}), with the pump-probe polarizations of horizontal-vertical (H-V), H-H, and V-V, respectively. Pronounced anisotropy is clearly seen in all the configurations.

\begin{figure*}
 \centering
  \includegraphics[width = 14.0 cm]{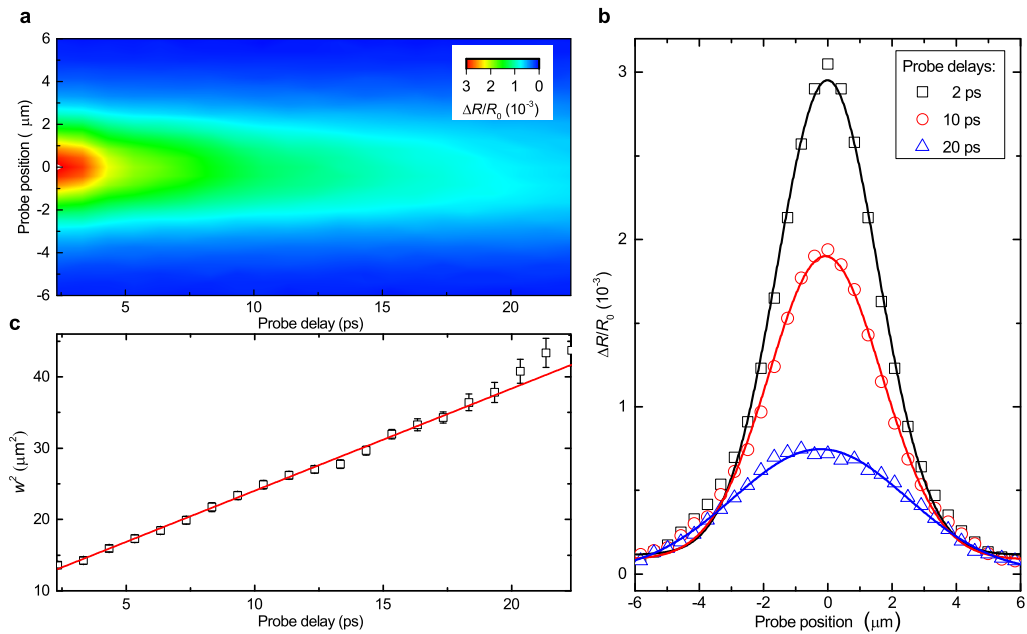}
  \caption{{\bf Photocarrier diffusion along the armchair direction of black phosphorus}. {\bf a}, Differential reflection as a function of the probe delay and probe position obtained by scanning the probe spot with respect to the pump spot along the armchair direction for each probe delay.  {\bf b}, Examples of the spatial profiles of the differential reflection signal, obtained with probe delays of 2 (squares), 10 (circles) and 20 ps (triangles), respectively. The solid lines are Gaussian fits.  {\bf c}, The squared width as a function of the probe delay. The solid line is a linear fit that gives a diffusion coefficient of 1,300 $\pm$ 30 cm$^2$s$^{-1}$.}
  \label{armchair}
\end{figure*}

The angular dependence of the differential reflection signal shown in Fig. 2{\bf b} can be induced by both the pump absorption and the probe detection efficiency. First, due to the different effective masses along the armchair and zigzag directions, the absorption coefficients are different for pump pulses polarized along the two directions. Hence, as the sample is rotated with respect to the pump polarization, the pump injects different carrier densities, which results in an angular dependent signal level. Second, the anisotropic bandstructure causes the probe pulses with different polarizations to sense the carriers with different efficiencies. Since the angular dependence in the H-V configuration is closer to H-H than V-V, we can conclude that the anisotropic differential reflection is mainly caused by the effect of pump absorption. When the pump polarization is along the long edge of the sample, a maximal carrier density is injected. Hence, we assign the long edge to the armchair direction, which is expected to have the maximal absorption\cite{nc54458}. At $\theta = 0$, the signal observed in the H-V configuration is about 1.7 times larger than H-H, indicating that the V-probe is more efficient in detecting carriers than the H-probe. That is, the maximal probe efficiency is achieved when the probe polarization is along the zigzag direction. 

The solid lines in Fig. 2{\bf b} are fits to the data by $A \mathrm{cos}^2(\theta + \theta _0) +B$. In the H-V configuration, we obtain a degree of anisotropy of $(A+B)/B = 6.7$. Given that the probe efficiency anisotropy is about 1.7, we deduce a pump absorption anisotropy of about a factor of 4. The pattern of H-H configuration is rotated with respect to H-V by about $8^\circ$, caused by a compromise of anisotropic pump and probe effects. The V-V pattern is rotated by $90^\circ$ from H-H, which confirms that the angular dependence observed is not from any artifacts of the experimental system.

\begin{figure*}
 \centering
  \includegraphics[width = 14.0 cm]{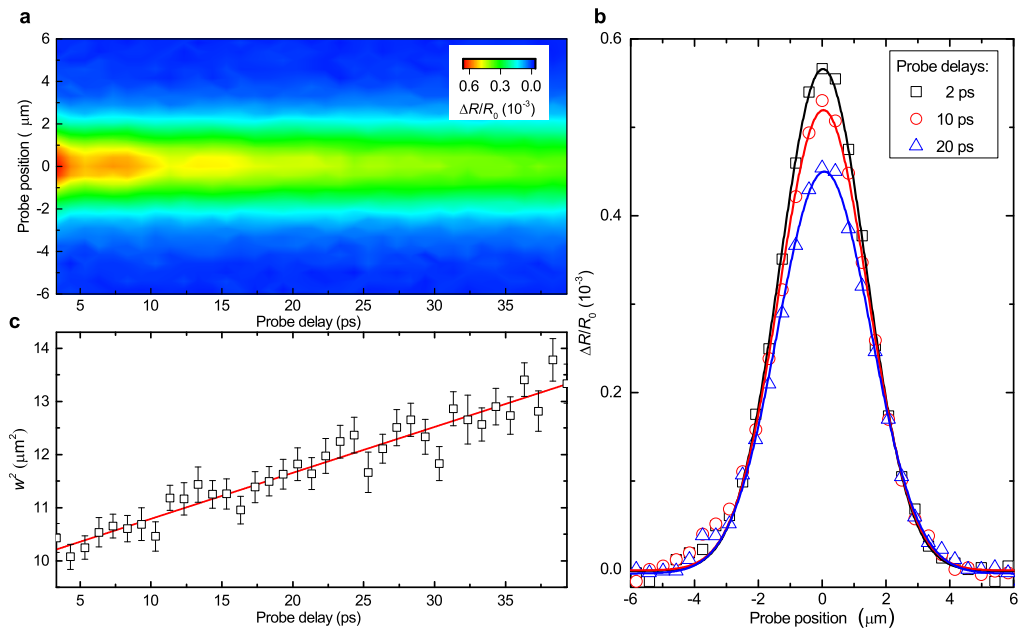}
  \caption{{\bf Photocarrier diffusion along the zigzag direction of black phosphorus}. {\bf a}, Differential reflection as a function of the probe delay and probe position obtained by scanning the probe spot with respect to the pump spot along the zigzag direction for each probe delay.  {\bf b}, Examples of the spatial profiles of the differential reflection signal, obtained with probe delays of 2 (squares), 10 (circles) and 20 ps (triangles), respectively. The solid lines are Gaussian fits.  {\bf c}, The squared width as a function of the probe delay. The solid line is a linear fit that gives a diffusion coefficient of 80 $\pm$ 5 cm$^2$s$^{-1}$.}
  \label{zigzag}
\end{figure*}

We study transport of photocarriers by performing spatially resolved differential reflection measurements. The tightly focused 730-nm pump injects photocarriers with a Gaussian spatial profile of about $w = 2.4~\mu$m in full width at half maximum. Once injected, the photocarriers diffuse in each atomic layer, driven by the in-plane density gradient. Hence, the spatial density profile expands. This diffusion process is monitored by measuring differential reflection of the probe as a function of the probe delay and probe position.

First, we study diffusion of photocarriers along the long edge direction of the flake, which is identified as the armchair direction according to Fig. 2{\bf b}. In this measurement, both pump and probe polarizations are along the armchair direction, and the probe spot is scanned along the same direction. Figure 3{\bf a} shows the differential reflection signal as a function of the probe delay and probe spot position with respect to the pump spot. At each probe delay, the profile has a Gaussian shape. Figure 3{\bf b} shows three examples of these profiles at probe delays of 2 (squares), 10 (circles), and 20 ps (triangles), respectively, along with the corresponding Gaussian fits (solid lines). From each fit, we obtain the width (full width at half maximum) of the profile. The squared width is plotted in Fig. 3{\bf c} as a function of the probe delay.

According to the diffusion model\cite{b385788}, the profile of carrier density remains its Gaussian shape during the diffusion, as confirmed in Fig. 3{\bf b}. The width of the profile evolves as $w^2 (t) = w^2 (t_0) + 16 \mathrm{ln}(2) D (t-t_0)$, where $t_0$ is an arbitrarily chosen initial time, and $D$ is the diffusion coefficient. Such a linear expansion of the area covered by the photocarriers is confirmed by the linear fit to the data shown in Fig. 3{\bf c} (red line). From the slope, we obtain a diffusion coefficient of  1,300 $\pm$ 30 cm$^2$s$^{-1}$.

Next, we rotate the sample by 90$^\circ$ and repeat the measurement under otherwise the same conditions. Now the polarizations and the scan directions are all along the zigzag direction. The results are presented in Fig. 4, in the same fashion as Fig. 3. Clearly, the diffusion is much slower along the zigzag direction. From a linear fit, shown as the solid line in Fig 4{\bf c}, we obtain a diffusion coefficient of 80 $\pm$ 5 cm$^2$s$^{-1}$, which is about a factor of 16 smaller than along the armchair direction.

The spatially resolved measurements also help clarify the fast transient process observed in Fig. 2{\bf a}, where the signal dropped by about a factor of 4 in the first 20 ps. As a rough estimation from Figs. 3{\bf c} and 4{\bf c}, the area of the profile increases by more than a factor of 2. Hence, the peak of the profile should decrease by the same factor due to the diffusion along. On the other hand, at later probe delays, the diffusion makes smaller contribution to the decay of the signal since the profile is already significantly broadened. This is confirmed by the single-exponential decay as indicated by the red line in Fig. 2{\bf a} and validates the attribution of the 100-ps time constant to the carrier lifetime.

Our noninvasive and all-optical measurement of spatiotemporal dynamics of photocarriers  in BP reveals its exceptional and highly anisotropic transport properties, which show the great potential of this material for electronic and optoelectronic applications. The measured diffusion coefficient can be used to deduce the mobility by using the Einstein's relation, $\mu / e = D/ k_B T$, where $\mu$, $e$, $k_B$, and $T$ are the mobility, elementary charge, Boltzmann constant, and temperature.  We obtain mobilities of 50,000 $\pm$ 2,000 and 3,000 $\pm$ 200 cm$^2$V$^{-1}$s$^{-1}$ along armchair and zigzag directions, respectively.

In the diffusion process we studied, the photocarriers move as electron-hole pairs due to the Coulomb attraction. Hence, the diffusion coefficient obtained are ambipolar diffusion coefficient, which is related to the diffusion coefficients of electrons and holes ($D_e$ and $D_h$) by $D_a = 2 D_e D_h / (D_e + D_h)$ for photocarriers\cite{Neamenbook}. In general, the ambipolar diffusion coefficient is controlled by the slower carriers. Hence, the ambipolar diffusion coefficient can be regarded as the lower limit of $D_e$ and D$_h$. In BP,  the effective masses of electrons and holes are within a factor of two\cite{icpsbp}. Hence, the ambipolar diffusion coefficient and ambipolar mobility provide satisfactory order-of-magnitude estimation of the diffusion coefficients and mobilities of both electrons and holes.

Recently, high room-temperature charge mobilities on the order of 10$^4$ cm$^2$V$^{-1}$s$^{-1}$ in BP has been predicted theoretically\cite{nc54475}. The previously reported experimental values have yet to reach that level. For example,  transistors of 10-nm BP flakes achieved mobilities in the range of 100 - 1000 cm$^2$V$^{-1}$s$^{-1}$ at room temperature\cite{acsnano84033,acsnano810035,nc54458}, and higher values at low temperatures\cite{nn9372,ieeeedl35795}. A trilayer BP transistor demonstrated electron and hole mobilities of 38 and 172 cm$^2$V$^{-1}$s$^{-1}$, respectively\cite{acsnano811730}. The electrical measurements may not reveal the intrinsic transport properties of the material since the transport can be limited by grain boundaries and by the quality of the electrodes. Indeed, significant influence of the metal contact has been suggested\cite{b90125441,sr46677,apl105163511}. On the other hand, the all-optical approach we used in this study does not involve device fabrication and grain boundaries play a negligible role in the diffusion process. We find that our results are reasonably consistent with the theoretical predictions\cite{nc54475}. 

The anisotropic transport property can be mainly attributed to the anisotropic effective masses of electrons and holes. The effective masses of electrons with wavevectors along the armchair and zigzag directions are $0.0825m_0$ and $1.027m_0$, respectively, where $m_0$ stands for the electron rest mass in vacuum. Hole effective masses are $0.0761m_0$ and $0.648m_0$ along the two directions, respectively\cite{icpsbp}. The differences in effective masses suggest that the mobilities along the armchair direction are about one order of magnitude larger than zigzag direction. Such a ratio is also predicted in monolayer and few-layer BP\cite{nc54475,jap116214505}. Based on our result, the mobility along the armchair direction is about 16 times larger than the zigzag direction, which is reasonably consistent with these predictions. 

The authors thank Jamie Wilt and Judy Wu for their help on the measurements of atomic force microscope. This material is based upon work supported by the National Science Foundation of USA (DMR-0954486, IIA-1430493), National Basic Research Program 973 of China (2011CB932700, 2011CB932703), Chinese Natural Science Fund Project (61335006, 61378073), and Fundamental Research Funds for the Central Universities (2015YJS181).


\begin{thebibliography}{36}%
\makeatletter
\providecommand \@ifxundefined [1]{%
 \@ifx{#1\undefined}
}%
\providecommand \@ifnum [1]{%
 \ifnum #1\expandafter \@firstoftwo
 \else \expandafter \@secondoftwo
 \fi
}%
\providecommand \@ifx [1]{%
 \ifx #1\expandafter \@firstoftwo
 \else \expandafter \@secondoftwo
 \fi
}%
\providecommand \natexlab [1]{#1}%
\providecommand \enquote  [1]{``#1''}%
\providecommand \bibnamefont  [1]{#1}%
\providecommand \bibfnamefont [1]{#1}%
\providecommand \citenamefont [1]{#1}%
\providecommand \href@noop [0]{\@secondoftwo}%
\providecommand \href [0]{\begingroup \@sanitize@url \@href}%
\providecommand \@href[1]{\@@startlink{#1}\@@href}%
\providecommand \@@href[1]{\endgroup#1\@@endlink}%
\providecommand \@sanitize@url [0]{\catcode `\\12\catcode `\$12\catcode
  `\&12\catcode `\#12\catcode `\^12\catcode `\_12\catcode `\%12\relax}%
\providecommand \@@startlink[1]{}%
\providecommand \@@endlink[0]{}%
\providecommand \url  [0]{\begingroup\@sanitize@url \@url }%
\providecommand \@url [1]{\endgroup\@href {#1}{\urlprefix }}%
\providecommand \urlprefix  [0]{URL }%
\providecommand \Eprint [0]{\href }%
\providecommand \doibase [0]{http://dx.doi.org/}%
\providecommand \selectlanguage [0]{\@gobble}%
\providecommand \bibinfo  [0]{\@secondoftwo}%
\providecommand \bibfield  [0]{\@secondoftwo}%
\providecommand \translation [1]{[#1]}%
\providecommand \BibitemOpen [0]{}%
\providecommand \bibitemStop [0]{}%
\providecommand \bibitemNoStop [0]{.\EOS\space}%
\providecommand \EOS [0]{\spacefactor3000\relax}%
\providecommand \BibitemShut  [1]{\csname bibitem#1\endcsname}%
\let\auto@bib@innerbib\@empty
\bibitem [{\citenamefont {Geim}\ and\ \citenamefont
  {Novoselov}(2007)}]{nm6183}%
  \BibitemOpen
  \bibfield  {author} {\bibinfo {author} {\bibfnamefont {A.~K.}\ \bibnamefont
  {Geim}}\ and\ \bibinfo {author} {\bibfnamefont {K.~S.}\ \bibnamefont
  {Novoselov}},\ }\href@noop {} {\bibfield  {journal} {\bibinfo  {journal}
  {Nat. Mater.}\ }\textbf {\bibinfo {volume} {6}},\ \bibinfo {pages} {183}
  (\bibinfo {year} {2007})}\BibitemShut {NoStop}%
\bibitem [{\citenamefont {Neto}\ \emph {et~al.}(2009)\citenamefont {Neto},
  \citenamefont {Guinea}, \citenamefont {Peres}, \citenamefont {Novoselov},\
  and\ \citenamefont {Geim}}]{rmp81109}%
  \BibitemOpen
  \bibfield  {author} {\bibinfo {author} {\bibfnamefont {A.~H.~C.}\
  \bibnamefont {Neto}}, \bibinfo {author} {\bibfnamefont {F.}~\bibnamefont
  {Guinea}}, \bibinfo {author} {\bibfnamefont {N.~M.~R.}\ \bibnamefont
  {Peres}}, \bibinfo {author} {\bibfnamefont {K.~S.}\ \bibnamefont
  {Novoselov}}, \ and\ \bibinfo {author} {\bibfnamefont {A.~K.}\ \bibnamefont
  {Geim}},\ }\href@noop {} {\bibfield  {journal} {\bibinfo  {journal} {Rev.
  Mod. Phys.}\ }\textbf {\bibinfo {volume} {81}},\ \bibinfo {pages} {109}
  (\bibinfo {year} {2009})}\BibitemShut {NoStop}%
\bibitem [{\citenamefont {Wang}\ \emph {et~al.}(2012)\citenamefont {Wang},
  \citenamefont {Kalantar-Zadeh}, \citenamefont {Kis}, \citenamefont
  {Coleman},\ and\ \citenamefont {Strano}}]{nn7699}%
  \BibitemOpen
  \bibfield  {author} {\bibinfo {author} {\bibfnamefont {Q.~H.}\ \bibnamefont
  {Wang}}, \bibinfo {author} {\bibfnamefont {K.}~\bibnamefont
  {Kalantar-Zadeh}}, \bibinfo {author} {\bibfnamefont {A.}~\bibnamefont {Kis}},
  \bibinfo {author} {\bibfnamefont {J.~N.}\ \bibnamefont {Coleman}}, \ and\
  \bibinfo {author} {\bibfnamefont {M.~S.}\ \bibnamefont {Strano}},\
  }\href@noop {} {\bibfield  {journal} {\bibinfo  {journal} {Nat.
  Nanotechnol.}\ }\textbf {\bibinfo {volume} {7}},\ \bibinfo {pages} {699}
  (\bibinfo {year} {2012})}\BibitemShut {NoStop}%
\bibitem [{\citenamefont {Mak}\ \emph {et~al.}(2010)\citenamefont {Mak},
  \citenamefont {Lee}, \citenamefont {Hone}, \citenamefont {Shan},\ and\
  \citenamefont {Heinz}}]{l105136805}%
  \BibitemOpen
  \bibfield  {author} {\bibinfo {author} {\bibfnamefont {K.~F.}\ \bibnamefont
  {Mak}}, \bibinfo {author} {\bibfnamefont {C.}~\bibnamefont {Lee}}, \bibinfo
  {author} {\bibfnamefont {J.}~\bibnamefont {Hone}}, \bibinfo {author}
  {\bibfnamefont {J.}~\bibnamefont {Shan}}, \ and\ \bibinfo {author}
  {\bibfnamefont {T.~F.}\ \bibnamefont {Heinz}},\ }\href@noop {} {\bibfield
  {journal} {\bibinfo  {journal} {Phys. Rev. Lett.}\ }\textbf {\bibinfo
  {volume} {105}},\ \bibinfo {pages} {136805} (\bibinfo {year}
  {2010})}\BibitemShut {NoStop}%
\bibitem [{\citenamefont {Splendiani}\ \emph {et~al.}(2010)\citenamefont
  {Splendiani}, \citenamefont {Sun}, \citenamefont {Zhang}, \citenamefont {Li},
  \citenamefont {Kim}, \citenamefont {Chim}, \citenamefont {Galli},\ and\
  \citenamefont {Wang}}]{nl101271}%
  \BibitemOpen
  \bibfield  {author} {\bibinfo {author} {\bibfnamefont {A.}~\bibnamefont
  {Splendiani}}, \bibinfo {author} {\bibfnamefont {L.}~\bibnamefont {Sun}},
  \bibinfo {author} {\bibfnamefont {Y.}~\bibnamefont {Zhang}}, \bibinfo
  {author} {\bibfnamefont {T.}~\bibnamefont {Li}}, \bibinfo {author}
  {\bibfnamefont {J.}~\bibnamefont {Kim}}, \bibinfo {author} {\bibfnamefont
  {C.~Y.}\ \bibnamefont {Chim}}, \bibinfo {author} {\bibfnamefont
  {G.}~\bibnamefont {Galli}}, \ and\ \bibinfo {author} {\bibfnamefont
  {F.}~\bibnamefont {Wang}},\ }\href@noop {} {\bibfield  {journal} {\bibinfo
  {journal} {Nano Lett.}\ }\textbf {\bibinfo {volume} {10}},\ \bibinfo {pages}
  {1271} (\bibinfo {year} {2010})}\BibitemShut {NoStop}%
\bibitem [{\citenamefont {Xiao}\ \emph {et~al.}(2012)\citenamefont {Xiao},
  \citenamefont {Liu}, \citenamefont {Feng}, \citenamefont {Xu},\ and\
  \citenamefont {Yao}}]{l108196802}%
  \BibitemOpen
  \bibfield  {author} {\bibinfo {author} {\bibfnamefont {D.}~\bibnamefont
  {Xiao}}, \bibinfo {author} {\bibfnamefont {G.~B.}\ \bibnamefont {Liu}},
  \bibinfo {author} {\bibfnamefont {W.}~\bibnamefont {Feng}}, \bibinfo {author}
  {\bibfnamefont {X.}~\bibnamefont {Xu}}, \ and\ \bibinfo {author}
  {\bibfnamefont {W.}~\bibnamefont {Yao}},\ }\href@noop {} {\bibfield
  {journal} {\bibinfo  {journal} {Phys. Rev. Lett.}\ }\textbf {\bibinfo
  {volume} {108}},\ \bibinfo {pages} {196802} (\bibinfo {year}
  {2012})}\BibitemShut {NoStop}%
\bibitem [{\citenamefont {Liu}\ \emph {et~al.}(2014{\natexlab{a}})\citenamefont
  {Liu}, \citenamefont {Neal}, \citenamefont {Zhu}, \citenamefont {Luo},
  \citenamefont {Xu}, \citenamefont {Tomanek},\ and\ \citenamefont
  {Ye}}]{acsnano84033}%
  \BibitemOpen
  \bibfield  {author} {\bibinfo {author} {\bibfnamefont {H.}~\bibnamefont
  {Liu}}, \bibinfo {author} {\bibfnamefont {A.~T.}\ \bibnamefont {Neal}},
  \bibinfo {author} {\bibfnamefont {Z.}~\bibnamefont {Zhu}}, \bibinfo {author}
  {\bibfnamefont {Z.}~\bibnamefont {Luo}}, \bibinfo {author} {\bibfnamefont
  {X.}~\bibnamefont {Xu}}, \bibinfo {author} {\bibfnamefont {D.}~\bibnamefont
  {Tomanek}}, \ and\ \bibinfo {author} {\bibfnamefont {P.~D.}\ \bibnamefont
  {Ye}},\ }\href@noop {} {\bibfield  {journal} {\bibinfo  {journal} {ACS Nano}\
  }\textbf {\bibinfo {volume} {8}},\ \bibinfo {pages} {4033} (\bibinfo {year}
  {2014}{\natexlab{a}})}\BibitemShut {NoStop}%
\bibitem [{\citenamefont {Brent}\ \emph {et~al.}(2014)\citenamefont {Brent},
  \citenamefont {Savjani}, \citenamefont {Lewis}, \citenamefont {Haigh},
  \citenamefont {Lewis},\ and\ \citenamefont {O'Brien}}]{cc5013338}%
  \BibitemOpen
  \bibfield  {author} {\bibinfo {author} {\bibfnamefont {J.~R.}\ \bibnamefont
  {Brent}}, \bibinfo {author} {\bibfnamefont {N.}~\bibnamefont {Savjani}},
  \bibinfo {author} {\bibfnamefont {E.~A.}\ \bibnamefont {Lewis}}, \bibinfo
  {author} {\bibfnamefont {S.~J.}\ \bibnamefont {Haigh}}, \bibinfo {author}
  {\bibfnamefont {D.~J.}\ \bibnamefont {Lewis}}, \ and\ \bibinfo {author}
  {\bibfnamefont {P.}~\bibnamefont {O'Brien}},\ }\href@noop {} {\bibfield
  {journal} {\bibinfo  {journal} {Chem. Commun.}\ }\textbf {\bibinfo {volume}
  {50}},\ \bibinfo {pages} {13338} (\bibinfo {year} {2014})}\BibitemShut
  {NoStop}%
\bibitem [{\citenamefont {Warschauer}(1963)}]{jap341853}%
  \BibitemOpen
  \bibfield  {author} {\bibinfo {author} {\bibfnamefont {D.}~\bibnamefont
  {Warschauer}},\ }\href@noop {} {\bibfield  {journal} {\bibinfo  {journal} {J.
  Appl. Phys.}\ }\textbf {\bibinfo {volume} {34}},\ \bibinfo {pages} {1853}
  (\bibinfo {year} {1963})}\BibitemShut {NoStop}%
\bibitem [{\citenamefont {Das}\ \emph {et~al.}(2014{\natexlab{a}})\citenamefont
  {Das}, \citenamefont {Zhang}, \citenamefont {Demarteau}, \citenamefont
  {Hoffmann}, \citenamefont {Dubey},\ and\ \citenamefont {Roelofs}}]{nl145733}%
  \BibitemOpen
  \bibfield  {author} {\bibinfo {author} {\bibfnamefont {S.}~\bibnamefont
  {Das}}, \bibinfo {author} {\bibfnamefont {W.}~\bibnamefont {Zhang}}, \bibinfo
  {author} {\bibfnamefont {M.}~\bibnamefont {Demarteau}}, \bibinfo {author}
  {\bibfnamefont {A.}~\bibnamefont {Hoffmann}}, \bibinfo {author}
  {\bibfnamefont {M.}~\bibnamefont {Dubey}}, \ and\ \bibinfo {author}
  {\bibfnamefont {A.}~\bibnamefont {Roelofs}},\ }\href@noop {} {\bibfield
  {journal} {\bibinfo  {journal} {Nano Lett.}\ }\textbf {\bibinfo {volume}
  {14}},\ \bibinfo {pages} {5733} (\bibinfo {year}
  {2014}{\natexlab{a}})}\BibitemShut {NoStop}%
\bibitem [{\citenamefont {Liang}\ \emph {et~al.}(2014)\citenamefont {Liang},
  \citenamefont {Wang}, \citenamefont {Lin}, \citenamefont {Sumpter},
  \citenamefont {Meunier},\ and\ \citenamefont {Pan}}]{nl146400}%
  \BibitemOpen
  \bibfield  {author} {\bibinfo {author} {\bibfnamefont {L.}~\bibnamefont
  {Liang}}, \bibinfo {author} {\bibfnamefont {J.}~\bibnamefont {Wang}},
  \bibinfo {author} {\bibfnamefont {W.}~\bibnamefont {Lin}}, \bibinfo {author}
  {\bibfnamefont {B.~G.}\ \bibnamefont {Sumpter}}, \bibinfo {author}
  {\bibfnamefont {V.}~\bibnamefont {Meunier}}, \ and\ \bibinfo {author}
  {\bibfnamefont {M.}~\bibnamefont {Pan}},\ }\href@noop {} {\bibfield
  {journal} {\bibinfo  {journal} {Nano Lett.}\ }\textbf {\bibinfo {volume}
  {14}},\ \bibinfo {pages} {6400} (\bibinfo {year} {2014})}\BibitemShut
  {NoStop}%
\bibitem [{\citenamefont {Zhang}\ \emph {et~al.}(2014)\citenamefont {Zhang},
  \citenamefont {Yang}, \citenamefont {Xu}, \citenamefont {Wang}, \citenamefont
  {Li}, \citenamefont {Ghufran}, \citenamefont {Zhang}, \citenamefont {Yu},
  \citenamefont {Zhang}, \citenamefont {Qin},\ and\ \citenamefont
  {Lu}}]{acsnano89590}%
  \BibitemOpen
  \bibfield  {author} {\bibinfo {author} {\bibfnamefont {S.}~\bibnamefont
  {Zhang}}, \bibinfo {author} {\bibfnamefont {J.}~\bibnamefont {Yang}},
  \bibinfo {author} {\bibfnamefont {R.}~\bibnamefont {Xu}}, \bibinfo {author}
  {\bibfnamefont {F.}~\bibnamefont {Wang}}, \bibinfo {author} {\bibfnamefont
  {W.}~\bibnamefont {Li}}, \bibinfo {author} {\bibfnamefont {M.}~\bibnamefont
  {Ghufran}}, \bibinfo {author} {\bibfnamefont {Y.~W.}\ \bibnamefont {Zhang}},
  \bibinfo {author} {\bibfnamefont {Z.}~\bibnamefont {Yu}}, \bibinfo {author}
  {\bibfnamefont {G.}~\bibnamefont {Zhang}}, \bibinfo {author} {\bibfnamefont
  {Q.}~\bibnamefont {Qin}}, \ and\ \bibinfo {author} {\bibfnamefont
  {Y.}~\bibnamefont {Lu}},\ }\href@noop {} {\bibfield  {journal} {\bibinfo
  {journal} {ACS Nano}\ }\textbf {\bibinfo {volume} {8}},\ \bibinfo {pages}
  {9590} (\bibinfo {year} {2014})}\BibitemShut {NoStop}%
\bibitem [{\citenamefont {Qiao}\ \emph {et~al.}(2014)\citenamefont {Qiao},
  \citenamefont {Kong}, \citenamefont {Hu}, \citenamefont {Yang},\ and\
  \citenamefont {Ji}}]{nc54475}%
  \BibitemOpen
  \bibfield  {author} {\bibinfo {author} {\bibfnamefont {J.}~\bibnamefont
  {Qiao}}, \bibinfo {author} {\bibfnamefont {X.}~\bibnamefont {Kong}}, \bibinfo
  {author} {\bibfnamefont {Z.~X.}\ \bibnamefont {Hu}}, \bibinfo {author}
  {\bibfnamefont {F.}~\bibnamefont {Yang}}, \ and\ \bibinfo {author}
  {\bibfnamefont {W.}~\bibnamefont {Ji}},\ }\href@noop {} {\bibfield  {journal}
  {\bibinfo  {journal} {Nat. Commun.}\ }\textbf {\bibinfo {volume} {5}},\
  \bibinfo {pages} {4475} (\bibinfo {year} {2014})}\BibitemShut {NoStop}%
\bibitem [{\citenamefont {Lam}(2014)}]{ieeeedl35963}%
  \BibitemOpen
  \bibfield  {author} {\bibinfo {author} {\bibfnamefont {K.~T.}\ \bibnamefont
  {Lam}},\ }\href@noop {} {\bibfield  {journal} {\bibinfo  {journal} {IEEE
  Elec. Dev. Lett.}\ }\textbf {\bibinfo {volume} {35}},\ \bibinfo {pages} {963}
  (\bibinfo {year} {2014})}\BibitemShut {NoStop}%
\bibitem [{\citenamefont {Liu}\ \emph {et~al.}(2014{\natexlab{b}})\citenamefont
  {Liu}, \citenamefont {Wang}, \citenamefont {Liu}, \citenamefont {Wang},\ and\
  \citenamefont {Guo}}]{ieeeted613871}%
  \BibitemOpen
  \bibfield  {author} {\bibinfo {author} {\bibfnamefont {F.}~\bibnamefont
  {Liu}}, \bibinfo {author} {\bibfnamefont {Y.~J.}\ \bibnamefont {Wang}},
  \bibinfo {author} {\bibfnamefont {X.~Y.}\ \bibnamefont {Liu}}, \bibinfo
  {author} {\bibfnamefont {J.}~\bibnamefont {Wang}}, \ and\ \bibinfo {author}
  {\bibfnamefont {H.}~\bibnamefont {Guo}},\ }\href@noop {} {\bibfield
  {journal} {\bibinfo  {journal} {IEEE T. Electron. Dev.}\ }\textbf {\bibinfo
  {volume} {61}},\ \bibinfo {pages} {3871} (\bibinfo {year}
  {2014}{\natexlab{b}})}\BibitemShut {NoStop}%
\bibitem [{\citenamefont {Du}\ \emph {et~al.}(2014)\citenamefont {Du},
  \citenamefont {Liu}, \citenamefont {Deng},\ and\ \citenamefont
  {Ye}}]{acsnano810035}%
  \BibitemOpen
  \bibfield  {author} {\bibinfo {author} {\bibfnamefont {Y.}~\bibnamefont
  {Du}}, \bibinfo {author} {\bibfnamefont {H.}~\bibnamefont {Liu}}, \bibinfo
  {author} {\bibfnamefont {Y.}~\bibnamefont {Deng}}, \ and\ \bibinfo {author}
  {\bibfnamefont {P.~D.}\ \bibnamefont {Ye}},\ }\href@noop {} {\bibfield
  {journal} {\bibinfo  {journal} {ACS Nano}\ }\textbf {\bibinfo {volume} {8}},\
  \bibinfo {pages} {10035} (\bibinfo {year} {2014})}\BibitemShut {NoStop}%
\bibitem [{\citenamefont {Xia}\ \emph {et~al.}(2014)\citenamefont {Xia},
  \citenamefont {Wang},\ and\ \citenamefont {Jia}}]{nc54458}%
  \BibitemOpen
  \bibfield  {author} {\bibinfo {author} {\bibfnamefont {F.}~\bibnamefont
  {Xia}}, \bibinfo {author} {\bibfnamefont {H.}~\bibnamefont {Wang}}, \ and\
  \bibinfo {author} {\bibfnamefont {Y.}~\bibnamefont {Jia}},\ }\href@noop {}
  {\bibfield  {journal} {\bibinfo  {journal} {Nat. Commun.}\ }\textbf {\bibinfo
  {volume} {5}},\ \bibinfo {pages} {4458} (\bibinfo {year} {2014})}\BibitemShut
  {NoStop}%
\bibitem [{\citenamefont {Li}\ \emph {et~al.}(2014)\citenamefont {Li},
  \citenamefont {Yu}, \citenamefont {Ye}, \citenamefont {Ge}, \citenamefont
  {Ou}, \citenamefont {Wu}, \citenamefont {Feng}, \citenamefont {Chen},\ and\
  \citenamefont {Zhang}}]{nn9372}%
  \BibitemOpen
  \bibfield  {author} {\bibinfo {author} {\bibfnamefont {L.}~\bibnamefont
  {Li}}, \bibinfo {author} {\bibfnamefont {Y.}~\bibnamefont {Yu}}, \bibinfo
  {author} {\bibfnamefont {G.~J.}\ \bibnamefont {Ye}}, \bibinfo {author}
  {\bibfnamefont {Q.}~\bibnamefont {Ge}}, \bibinfo {author} {\bibfnamefont
  {X.}~\bibnamefont {Ou}}, \bibinfo {author} {\bibfnamefont {H.}~\bibnamefont
  {Wu}}, \bibinfo {author} {\bibfnamefont {D.}~\bibnamefont {Feng}}, \bibinfo
  {author} {\bibfnamefont {X.~H.}\ \bibnamefont {Chen}}, \ and\ \bibinfo
  {author} {\bibfnamefont {Y.}~\bibnamefont {Zhang}},\ }\href@noop {}
  {\bibfield  {journal} {\bibinfo  {journal} {Nat. Nanotechnol.}\ }\textbf
  {\bibinfo {volume} {9}},\ \bibinfo {pages} {372} (\bibinfo {year}
  {2014})}\BibitemShut {NoStop}%
\bibitem [{\citenamefont {Liu}\ \emph {et~al.}(2014{\natexlab{c}})\citenamefont
  {Liu}, \citenamefont {Neal}, \citenamefont {Si}, \citenamefont {Du},\ and\
  \citenamefont {Ye}}]{ieeeedl35795}%
  \BibitemOpen
  \bibfield  {author} {\bibinfo {author} {\bibfnamefont {H.}~\bibnamefont
  {Liu}}, \bibinfo {author} {\bibfnamefont {A.~T.}\ \bibnamefont {Neal}},
  \bibinfo {author} {\bibfnamefont {M.~W.}\ \bibnamefont {Si}}, \bibinfo
  {author} {\bibfnamefont {Y.~C.}\ \bibnamefont {Du}}, \ and\ \bibinfo {author}
  {\bibfnamefont {P.~D.}\ \bibnamefont {Ye}},\ }\href@noop {} {\bibfield
  {journal} {\bibinfo  {journal} {IEEE Electr. Dev. Lett.}\ }\textbf {\bibinfo
  {volume} {35}},\ \bibinfo {pages} {795} (\bibinfo {year}
  {2014}{\natexlab{c}})}\BibitemShut {NoStop}%
\bibitem [{\citenamefont {Das}\ \emph {et~al.}(2014{\natexlab{b}})\citenamefont
  {Das}, \citenamefont {Demarteau},\ and\ \citenamefont
  {Roelofs}}]{acsnano811730}%
  \BibitemOpen
  \bibfield  {author} {\bibinfo {author} {\bibfnamefont {S.}~\bibnamefont
  {Das}}, \bibinfo {author} {\bibfnamefont {M.}~\bibnamefont {Demarteau}}, \
  and\ \bibinfo {author} {\bibfnamefont {A.}~\bibnamefont {Roelofs}},\
  }\href@noop {} {\bibfield  {journal} {\bibinfo  {journal} {ACS Nano}\
  }\textbf {\bibinfo {volume} {8}},\ \bibinfo {pages} {11730} (\bibinfo {year}
  {2014}{\natexlab{b}})}\BibitemShut {NoStop}%
\bibitem [{\citenamefont {Na}\ \emph {et~al.}(2014)\citenamefont {Na},
  \citenamefont {Lee}, \citenamefont {Lim}, \citenamefont {do}, \citenamefont
  {Kim}, \citenamefont {Choi},\ and\ \citenamefont {Song}}]{acsnano811753}%
  \BibitemOpen
  \bibfield  {author} {\bibinfo {author} {\bibfnamefont {J.}~\bibnamefont
  {Na}}, \bibinfo {author} {\bibfnamefont {Y.~T.}\ \bibnamefont {Lee}},
  \bibinfo {author} {\bibfnamefont {J.~A.}\ \bibnamefont {Lim}}, \bibinfo
  {author} {\bibfnamefont {K.~H.}\ \bibnamefont {do}}, \bibinfo {author}
  {\bibfnamefont {G.~T.}\ \bibnamefont {Kim}}, \bibinfo {author} {\bibfnamefont
  {W.~K.}\ \bibnamefont {Choi}}, \ and\ \bibinfo {author} {\bibfnamefont
  {Y.~W.}\ \bibnamefont {Song}},\ }\href@noop {} {\bibfield  {journal}
  {\bibinfo  {journal} {ACS Nano}\ }\textbf {\bibinfo {volume} {8}},\ \bibinfo
  {pages} {11753} (\bibinfo {year} {2014})}\BibitemShut {NoStop}%
\bibitem [{\citenamefont {Wood}\ \emph {et~al.}(2014)\citenamefont {Wood},
  \citenamefont {Wells}, \citenamefont {Jariwala}, \citenamefont {Chen},
  \citenamefont {Cho}, \citenamefont {Sangwan}, \citenamefont {Liu},
  \citenamefont {Lauhon}, \citenamefont {Marks},\ and\ \citenamefont
  {Hersam}}]{nl146964}%
  \BibitemOpen
  \bibfield  {author} {\bibinfo {author} {\bibfnamefont {J.~D.}\ \bibnamefont
  {Wood}}, \bibinfo {author} {\bibfnamefont {S.~A.}\ \bibnamefont {Wells}},
  \bibinfo {author} {\bibfnamefont {D.}~\bibnamefont {Jariwala}}, \bibinfo
  {author} {\bibfnamefont {K.~S.}\ \bibnamefont {Chen}}, \bibinfo {author}
  {\bibfnamefont {E.}~\bibnamefont {Cho}}, \bibinfo {author} {\bibfnamefont
  {V.~K.}\ \bibnamefont {Sangwan}}, \bibinfo {author} {\bibfnamefont
  {X.}~\bibnamefont {Liu}}, \bibinfo {author} {\bibfnamefont {L.~J.}\
  \bibnamefont {Lauhon}}, \bibinfo {author} {\bibfnamefont {T.~J.}\
  \bibnamefont {Marks}}, \ and\ \bibinfo {author} {\bibfnamefont {M.~C.}\
  \bibnamefont {Hersam}},\ }\href@noop {} {\bibfield  {journal} {\bibinfo
  {journal} {Nano Lett.}\ }\textbf {\bibinfo {volume} {14}},\ \bibinfo {pages}
  {6964} (\bibinfo {year} {2014})}\BibitemShut {NoStop}%
\bibitem [{\citenamefont {Wang}\ \emph {et~al.}(2014)\citenamefont {Wang},
  \citenamefont {Wang}, \citenamefont {Xia}, \citenamefont {Wang},
  \citenamefont {Jiang}, \citenamefont {Xia}, \citenamefont {Chin},
  \citenamefont {Dubey},\ and\ \citenamefont {Han}}]{nl146424}%
  \BibitemOpen
  \bibfield  {author} {\bibinfo {author} {\bibfnamefont {H.}~\bibnamefont
  {Wang}}, \bibinfo {author} {\bibfnamefont {X.}~\bibnamefont {Wang}}, \bibinfo
  {author} {\bibfnamefont {F.}~\bibnamefont {Xia}}, \bibinfo {author}
  {\bibfnamefont {L.}~\bibnamefont {Wang}}, \bibinfo {author} {\bibfnamefont
  {H.}~\bibnamefont {Jiang}}, \bibinfo {author} {\bibfnamefont
  {Q.}~\bibnamefont {Xia}}, \bibinfo {author} {\bibfnamefont {M.~L.}\
  \bibnamefont {Chin}}, \bibinfo {author} {\bibfnamefont {M.}~\bibnamefont
  {Dubey}}, \ and\ \bibinfo {author} {\bibfnamefont {S.~J.}\ \bibnamefont
  {Han}},\ }\href@noop {} {\bibfield  {journal} {\bibinfo  {journal} {Nano
  Lett.}\ }\textbf {\bibinfo {volume} {14}},\ \bibinfo {pages} {6424} (\bibinfo
  {year} {2014})}\BibitemShut {NoStop}%
\bibitem [{\citenamefont {Buscema}\ \emph
  {et~al.}(2014{\natexlab{a}})\citenamefont {Buscema}, \citenamefont
  {Groenendijk}, \citenamefont {Blanter}, \citenamefont {Steele}, \citenamefont
  {van~der Zant},\ and\ \citenamefont {Castellanos-Gomez}}]{nl143347}%
  \BibitemOpen
  \bibfield  {author} {\bibinfo {author} {\bibfnamefont {M.}~\bibnamefont
  {Buscema}}, \bibinfo {author} {\bibfnamefont {D.~J.}\ \bibnamefont
  {Groenendijk}}, \bibinfo {author} {\bibfnamefont {S.~I.}\ \bibnamefont
  {Blanter}}, \bibinfo {author} {\bibfnamefont {G.~A.}\ \bibnamefont {Steele}},
  \bibinfo {author} {\bibfnamefont {H.~S.}\ \bibnamefont {van~der Zant}}, \
  and\ \bibinfo {author} {\bibfnamefont {A.}~\bibnamefont
  {Castellanos-Gomez}},\ }\href@noop {} {\bibfield  {journal} {\bibinfo
  {journal} {Nano Lett.}\ }\textbf {\bibinfo {volume} {14}},\ \bibinfo {pages}
  {3347} (\bibinfo {year} {2014}{\natexlab{a}})}\BibitemShut {NoStop}%
\bibitem [{\citenamefont {Engel}\ \emph {et~al.}(2014)\citenamefont {Engel},
  \citenamefont {Steiner},\ and\ \citenamefont {Avouris}}]{nl146414}%
  \BibitemOpen
  \bibfield  {author} {\bibinfo {author} {\bibfnamefont {M.}~\bibnamefont
  {Engel}}, \bibinfo {author} {\bibfnamefont {M.}~\bibnamefont {Steiner}}, \
  and\ \bibinfo {author} {\bibfnamefont {P.}~\bibnamefont {Avouris}},\
  }\href@noop {} {\bibfield  {journal} {\bibinfo  {journal} {Nano Lett.}\
  }\textbf {\bibinfo {volume} {14}},\ \bibinfo {pages} {6414} (\bibinfo {year}
  {2014})}\BibitemShut {NoStop}%
\bibitem [{\citenamefont {Buscema}\ \emph
  {et~al.}(2014{\natexlab{b}})\citenamefont {Buscema}, \citenamefont
  {Groenendijk}, \citenamefont {Steele}, \citenamefont {van~der Zant},\ and\
  \citenamefont {Castellanos-Gomez}}]{nc54651}%
  \BibitemOpen
  \bibfield  {author} {\bibinfo {author} {\bibfnamefont {M.}~\bibnamefont
  {Buscema}}, \bibinfo {author} {\bibfnamefont {D.~J.}\ \bibnamefont
  {Groenendijk}}, \bibinfo {author} {\bibfnamefont {G.~A.}\ \bibnamefont
  {Steele}}, \bibinfo {author} {\bibfnamefont {H.~S.}\ \bibnamefont {van~der
  Zant}}, \ and\ \bibinfo {author} {\bibfnamefont {A.}~\bibnamefont
  {Castellanos-Gomez}},\ }\href@noop {} {\bibfield  {journal} {\bibinfo
  {journal} {Nat. Commun.}\ }\textbf {\bibinfo {volume} {5}},\ \bibinfo {pages}
  {4651} (\bibinfo {year} {2014}{\natexlab{b}})}\BibitemShut {NoStop}%
\bibitem [{\citenamefont {Ong}\ \emph {et~al.}(2014)\citenamefont {Ong},
  \citenamefont {Zhang},\ and\ \citenamefont {Zhang}}]{jap116214505}%
  \BibitemOpen
  \bibfield  {author} {\bibinfo {author} {\bibfnamefont {Z.~Y.}\ \bibnamefont
  {Ong}}, \bibinfo {author} {\bibfnamefont {G.}~\bibnamefont {Zhang}}, \ and\
  \bibinfo {author} {\bibfnamefont {Y.~W.}\ \bibnamefont {Zhang}},\ }\href@noop
  {} {\bibfield  {journal} {\bibinfo  {journal} {J. Appl. Phys.}\ }\textbf
  {\bibinfo {volume} {116}},\ \bibinfo {pages} {214505} (\bibinfo {year}
  {2014})}\BibitemShut {NoStop}%
\bibitem [{\citenamefont {Fei}\ and\ \citenamefont {Yang}(2014)}]{nl142884}%
  \BibitemOpen
  \bibfield  {author} {\bibinfo {author} {\bibfnamefont {R.}~\bibnamefont
  {Fei}}\ and\ \bibinfo {author} {\bibfnamefont {L.}~\bibnamefont {Yang}},\
  }\href@noop {} {\bibfield  {journal} {\bibinfo  {journal} {Nano Lett.}\
  }\textbf {\bibinfo {volume} {14}},\ \bibinfo {pages} {2884} (\bibinfo {year}
  {2014})}\BibitemShut {NoStop}%
\bibitem [{\citenamefont {Li}\ and\ \citenamefont
  {Appelbaum}(2014)}]{b90115439}%
  \BibitemOpen
  \bibfield  {author} {\bibinfo {author} {\bibfnamefont {P.}~\bibnamefont
  {Li}}\ and\ \bibinfo {author} {\bibfnamefont {I.}~\bibnamefont {Appelbaum}},\
  }\href@noop {} {\bibfield  {journal} {\bibinfo  {journal} {Phys. Rev. B}\
  }\textbf {\bibinfo {volume} {90}},\ \bibinfo {pages} {115439} (\bibinfo
  {year} {2014})}\BibitemShut {NoStop}%
\bibitem [{\citenamefont {Low}\ \emph {et~al.}(2014)\citenamefont {Low},
  \citenamefont {Rodin}, \citenamefont {Carvalho}, \citenamefont {Jiang},
  \citenamefont {Wang}, \citenamefont {Xia},\ and\ \citenamefont
  {Neto}}]{b90075434}%
  \BibitemOpen
  \bibfield  {author} {\bibinfo {author} {\bibfnamefont {T.}~\bibnamefont
  {Low}}, \bibinfo {author} {\bibfnamefont {A.~S.}\ \bibnamefont {Rodin}},
  \bibinfo {author} {\bibfnamefont {A.}~\bibnamefont {Carvalho}}, \bibinfo
  {author} {\bibfnamefont {Y.~J.}\ \bibnamefont {Jiang}}, \bibinfo {author}
  {\bibfnamefont {H.}~\bibnamefont {Wang}}, \bibinfo {author} {\bibfnamefont
  {F.~N.}\ \bibnamefont {Xia}}, \ and\ \bibinfo {author} {\bibfnamefont
  {A.~H.~C.}\ \bibnamefont {Neto}},\ }\href@noop {} {\bibfield  {journal}
  {\bibinfo  {journal} {Phys. Rev. B}\ }\textbf {\bibinfo {volume} {90}},\
  \bibinfo {pages} {075434} (\bibinfo {year} {2014})}\BibitemShut {NoStop}%
\bibitem [{\citenamefont {Smith}\ \emph {et~al.}(1988)\citenamefont {Smith},
  \citenamefont {Wake}, \citenamefont {Wolfe}, \citenamefont {Levi},
  \citenamefont {Klein}, \citenamefont {Klem}, \citenamefont {Henderson},\ and\
  \citenamefont {Morko{\c c}}}]{b385788}%
  \BibitemOpen
  \bibfield  {author} {\bibinfo {author} {\bibfnamefont {L.~M.}\ \bibnamefont
  {Smith}}, \bibinfo {author} {\bibfnamefont {D.~R.}\ \bibnamefont {Wake}},
  \bibinfo {author} {\bibfnamefont {J.~P.}\ \bibnamefont {Wolfe}}, \bibinfo
  {author} {\bibfnamefont {D.}~\bibnamefont {Levi}}, \bibinfo {author}
  {\bibfnamefont {M.~V.}\ \bibnamefont {Klein}}, \bibinfo {author}
  {\bibfnamefont {J.}~\bibnamefont {Klem}}, \bibinfo {author} {\bibfnamefont
  {T.}~\bibnamefont {Henderson}}, \ and\ \bibinfo {author} {\bibfnamefont
  {H.}~\bibnamefont {Morko{\c c}}},\ }\href@noop {} {\bibfield  {journal}
  {\bibinfo  {journal} {Phys.~Rev.~B}\ }\textbf {\bibinfo {volume} {38}},\
  \bibinfo {pages} {5788} (\bibinfo {year} {1988})}\BibitemShut {NoStop}%
\bibitem [{\citenamefont {Neamen}(2002)}]{Neamenbook}%
  \BibitemOpen
  \bibfield  {author} {\bibinfo {author} {\bibfnamefont {D.~A.}\ \bibnamefont
  {Neamen}},\ }\href@noop {} {\emph {\bibinfo {title} {Semiconductor physics
  and devices}}}\ (\bibinfo  {publisher} {McGraw-Hill, Boston},\ \bibinfo
  {year} {2002})\BibitemShut {NoStop}%
\bibitem [{\citenamefont {Morita}\ \emph {et~al.}(1985)\citenamefont {Morita},
  \citenamefont {Asahina}, \citenamefont {Kaneta},\ and\ \citenamefont
  {Sasaki}}]{icpsbp}%
  \BibitemOpen
  \bibfield  {author} {\bibinfo {author} {\bibfnamefont {A.}~\bibnamefont
  {Morita}}, \bibinfo {author} {\bibfnamefont {H.}~\bibnamefont {Asahina}},
  \bibinfo {author} {\bibfnamefont {C.}~\bibnamefont {Kaneta}}, \ and\ \bibinfo
  {author} {\bibfnamefont {T.}~\bibnamefont {Sasaki}},\ }\href@noop {}
  {\bibfield  {journal} {\bibinfo  {journal} {Proc. 17th Int. Conf. Phys.
  Semocond.}\ ,\ \bibinfo {pages} {1320}} (\bibinfo {year} {1985})}\BibitemShut
  {NoStop}%
\bibitem [{\citenamefont {Gong}\ \emph {et~al.}(2014)\citenamefont {Gong},
  \citenamefont {Zhang}, \citenamefont {Ji},\ and\ \citenamefont
  {Guo}}]{b90125441}%
  \BibitemOpen
  \bibfield  {author} {\bibinfo {author} {\bibfnamefont {K.}~\bibnamefont
  {Gong}}, \bibinfo {author} {\bibfnamefont {L.}~\bibnamefont {Zhang}},
  \bibinfo {author} {\bibfnamefont {W.}~\bibnamefont {Ji}}, \ and\ \bibinfo
  {author} {\bibfnamefont {H.}~\bibnamefont {Guo}},\ }\href@noop {} {\bibfield
  {journal} {\bibinfo  {journal} {Phys. Rev. B}\ }\textbf {\bibinfo {volume}
  {90}},\ \bibinfo {pages} {125441} (\bibinfo {year} {2014})}\BibitemShut
  {NoStop}%
\bibitem [{\citenamefont {Cai}\ \emph {et~al.}(2014)\citenamefont {Cai},
  \citenamefont {Zhang},\ and\ \citenamefont {Zhang}}]{sr46677}%
  \BibitemOpen
  \bibfield  {author} {\bibinfo {author} {\bibfnamefont {Y.}~\bibnamefont
  {Cai}}, \bibinfo {author} {\bibfnamefont {G.}~\bibnamefont {Zhang}}, \ and\
  \bibinfo {author} {\bibfnamefont {Y.~W.}\ \bibnamefont {Zhang}},\ }\href@noop
  {} {\bibfield  {journal} {\bibinfo  {journal} {Sci. Rep.}\ }\textbf {\bibinfo
  {volume} {4}},\ \bibinfo {pages} {6677} (\bibinfo {year} {2014})}\BibitemShut
  {NoStop}%
\bibitem [{\citenamefont {Wan}\ \emph {et~al.}(2014)\citenamefont {Wan},
  \citenamefont {Cao},\ and\ \citenamefont {Guo}}]{apl105163511}%
  \BibitemOpen
  \bibfield  {author} {\bibinfo {author} {\bibfnamefont {R.}~\bibnamefont
  {Wan}}, \bibinfo {author} {\bibfnamefont {X.}~\bibnamefont {Cao}}, \ and\
  \bibinfo {author} {\bibfnamefont {J.}~\bibnamefont {Guo}},\ }\href@noop {}
  {\bibfield  {journal} {\bibinfo  {journal} {Appl. Phys. Lett.}\ }\textbf
  {\bibinfo {volume} {105}},\ \bibinfo {pages} {163511} (\bibinfo {year}
  {2014})}\BibitemShut {NoStop}%
\end{thebibliography}

%

\end{document}